\documentclass[a4paper, 10pt]{article}
\usepackage[utf8x]{inputenc}
\usepackage[english]{babel}
\usepackage{amsfonts, amssymb, amsmath, amsthm, stmaryrd, ytableau,
 cancel, mathrsfs, mathcomp, calrsfs,ulem}
\usepackage{pifont, indentfirst, dsfont, relsize}
\usepackage{geometry, graphicx, tabularx,cite}
\usepackage{fancyhdr}
\usepackage{color}
\definecolor{linkcolor}{rgb}{0,0,0.6}
\usepackage[	pdftex,colorlinks=true,	
			pdfstartview=FitV,
			linkcolor= linkcolor,
			citecolor= linkcolor,
			urlcolor= linkcolor,
			hyperindex=true,
			hyperfigures=false]
			{hyperref}

\usepackage{lastpage}
\date{}
\definecolor{rougef}{rgb}{0.56,0,0}
\definecolor{vertf}{rgb}{0,0.5,0}
\definecolor{bleuf}{rgb}{0,0,0.8}

\begin{document}

\title{\textbf{Higher-derivative harmonic oscillators: stability of classical dynamics and adiabatic invariants}}

\author{Nicolas Boulanger$^1$ \and Fabien Buisseret$^{2,3}$ \and Fr\'{e}d\'{e}ric Dierick$^{2,4}$ \and Olivier White$^{5}$}

\date{$^1$ Service de Physique de l'Univers, Champs et Gravitation, Universit\'{e} de Mons, UMONS  Research
Institute for Complex Systems, Place du Parc 20, 7000 Mons, Belgium \\
$^2$ Forme $\&$ Fonctionnement Humain Lab, Department of Physical Therapy, CERISIC, Haute Ecole Louvain en Hainaut, 136 rue Trieu Kaisin, 6061 Montignies sur Sambre, Belgium \\
$^3$ Service de Physique Nucl\'{e}aire et Subnucl\'{e}aire, Universit\'{e} de Mons, UMONS Research Institute for Complex Systems, 20 Place du Parc, 7000 Mons, Belgium \\
$^4$ Facult\'{e} des Sciences de la Motricit\'e, Universit\'e catholique de Louvain, 1 Place Pierre de Coubertin, 1348 Louvain-la-Neuve, Belgium \\
$^5$ Universit\'{e} de Bourgogne INSERM-U1093 Cognition, Action, and Sensorimotor Plasticity, Campus Universitaire, BP 27877, 21078 Dijon, France\\[6ex]}
\maketitle
\begin{abstract}
    The status of classical stability in higher-derivative  systems is still
    subject to discussions. In this note, we argue that, contrary to general belief, 
    many higher-derivative systems are classically stable. The main tool to see
    this property are Nekhoroshev's estimates relying on the action-angle
    formulation of classical mechanics. The latter formulation can be reached provided the Hamiltonian is separable, which is the case for higher-derivative harmonic oscillators. The Pais-Uhlenbeck oscillators appear to be the only type of higher-derivative 
    harmonic oscillator with stable classical dynamics. A wide class of interaction 
    potentials can even be added that preserve classical stability. 
    Adiabatic invariants are built in the case of a Pais-Uhlenbeck oscillator 
    slowly changing in time; it is shown indeed that the dynamical stability is 
    not jeopardised by the time-dependent perturbation.
\end{abstract}

\thispagestyle{empty}
%\tableofcontents
\newpage

\setcounter{page}{1}

%*******************************************
\section{Introduction}\label{sec:Intro}
%*******************************************

Variational principles based on action functionals of the form $S[x]=\int L(x,\dot x)\, dt$ 
have a special status in the sense that Newtonian mechanics can be recovered from 
Lagrangians $L(x,\dot x)$ depending only on position and velocity. 
Still, Lagrangians functions depending on higher time derivatives of the 
position, \textit{i.e}, higher-derivative (HD) Lagrangians, are also worth of interest. 
Let us mention three areas in which HD models are encountered: 
\begin{enumerate}
    \item Explicit construction of classicaly stable and unstable HD dynamical systems:
    A class of classical HD harmonic oscillators was proposed in Sec.~II 
    of \cite{Pais:1950za}
    and is still actively studied nowadays under the name of Pais-Uhlenbeck (P-U) oscillator, 
    see \textit{e.g.} Refs \cite{Nesterenko:2006tu,Smilga:2008pr,Joung:2012sa,Pavsic:2013mja,Kaparulin:2014vpa,Woodard:2015zca,Masterov:2016jft,Avendano-Camacho:2017ykt,Smilga:2017arl,Nicolis:2018lkg,Horvathy:2002vt,Plyushchay:2006pw} and references therein for recent contributions to the field;
    \item Renormalizability of HD field theories:
    In their pioneering work \cite{Pais:1950za}, Pais and Uhlenbeck addressed the issue of renormalizability in field theory through the inclusion of HD terms. 
    HD gravities, like Weyl gravity, are promising renormalisable models of quantum gravity, see the seminal paper \cite{Stelle:1976gc} and recent references 
    in \cite{Nitta:2018yzb,Giribet:2018hck}. 
    These HD models bring in the Einstein-Hilbert term upon radiative 
    corrections, see e.g. \cite{Tomboulis:1980bs}. 
    They are also interesting in the context of cosmology
    and supergravity, 
    see \cite{Schmidt:1994iz,Hawking:2001yt,Ferrara:2018wqd,Castellanos:2018dub} and Refs. therein;
    \item HD effective dynamics of voluntary human motions:
    The underlying dynamics of such motions is expected to involve HD variational principles such as minimal jerk, see \textit{e.g.} Refs \cite{motorC}. 
    In this case, higher derivative terms may be thought of as a way to account for 
    an intrinsic nonlocality (in time) of planified motion: Motor control may indeed add memory effects to standard Newtonian dynamics, which 
    may be translated into a HD effective action.
\end{enumerate}

A key feature of classical HD dynamics is that the energy has no definite sign, as it is 
readily observed from the general structure of HD Hamiltonians \cite{Ostro}. 
The presence of HD terms may lead to unbounded trajectories at the classical level  -- an 
explicit case is built in \cite{Nesterenko:2006tu} -- and to loss of unitarity 
at the quantum level \cite{Smilga:2017arl}. However several cases are known for which 
classical trajectories are bounded and unitarity is preserved at the quantum 
level \cite{Smilga:2017arl}. Having these recent results in mind, we think that 
providing a general method in order to assess classical stability of HD models 
is worth of interest. This is the main goal of the present paper, in which we 
focus on the case of HD harmonic oscillators. 
In the case of perturbed harmonic oscillators, an important body of works
concerning their stability has been
produced \cite{benett, valdi} that seems to have gone unnoticed by 
the HD community, so far. In a very specific case of a P-U oscillator with 
at most two time derivatives in the Lagrangian, Pagani et al. proved stability 
under a general class of cubic and quartic interactions \cite{pagani}.

The present paper is organised as follows. 
The Lagrangian (Sec. \ref{sec:lag}) and Hamiltonian (Sec. \ref{sec:ham}) formulations of HD 
harmonic oscillators are reviewed and the necessary conditions 
for their classical trajectories to be bounded are established.
The dynamics is then formulated in terms of the action-angle coordinates in Sec. \ref{sec:ac-an}  and adiabatic invariants are computed. 
Finally, classical stability against time-dependent perturbations is discussed 
by using Nekhoroshev's estimates \cite{nekho}.

%*******************************************
\section{Lagrangian formulation}\label{sec:lag}
%*******************************************

In this section we review the Lagrangian formulation of HD classical systems 
with finitely many degrees of freedom, in essentially the way that was presented 
long ago by Ostrogradki \cite{Ostro}. We then review the Pais-Uhlenbeck 
parametrisation \cite{Pais:1950za} of HD Lagrangians. 

\subsection{Generalities}\label{sec:gene}

Let $L(x^{(0)},x^{(1)},\ldots , x^{(N)})$ be a Lagrangian depending on 
the $N$ first derivatives of the dynamical variables $x(t):=x^{(0)}(t)$, $x^{(1)}(t):=\dot x(t)$ etc. 
possibly upon adding total derivatives in order to lower as much as 
possible the order of derivatives of $x(t)\,$. 
The action, evaluated between time $t_1$ and time 
$t_2\,$, is $S[x]=\int_{t_1}^{t_2} L\, dt\,$.
Hamilton's variational principle $\delta S=0$ implies 
the equations of motion
\begin{equation}
0 = \frac{\delta L}{\delta x} \equiv \sum^N_{j=0} \left(-\frac{d}{dt}\right)^j\, 
\frac{\partial L}{\partial x^{(j)}} \;,
\label{eom}
\end{equation}
together with the vanishing of the boundary terms
\begin{eqnarray}
    \left. \sum_{i=0}^{N-1} \delta x^{(i)}\,p_{i}\right|^{t_2}_{t_1}\;\; &=& 0\;,
    \nonumber \\
    p_{i}&:=& \frac{\delta L}{\delta x^{(i+1)}}
    \equiv \sum^{N-i-1}_{j=0} \left(-\frac{d}{dt}\right)^j\, 
\frac{\partial L}{\partial x^{(i+j+1)}}
    \;, \quad i\in\{0,\ldots,N-1\}\;.
    \label{momenta}
\end{eqnarray}
One chooses to cancel the above boundary terms by imposing the following 
conditions at the boundaries of the integration domain:
\begin{equation}
    \delta x^{(i)}(t_2) = 0 = \delta x^{(i)}(t_1)\;, \quad \forall ~i\in\{0,\ldots,N-1\}\;. \nonumber
\end{equation}
It amounts to declaring that the initial data needed for solving the equation 
(\ref{eom}) is 
given by the values of $x^{(0)}\,$, $x^{(1)}\,$, $\ldots\,$, $x^{(N-2)}$ and 
$x^{(2N-1)}$ at initial time $t_1\,$. 
Indeed, provided one assumes the regularity condition
\begin{equation}
\frac{\partial^2 L}{\partial x^{(N)}\partial x^{(N)}} \neq 0\;,    \nonumber
\label{regularity}
\end{equation}
the above initial data give a well-posed Cauchy problem for the ordinary differential
equation (\ref{eom}).

\subsection{Toy model}

The simple HD harmonic oscillator
\begin{equation}\label{ltoy}
L_{{\rm toy}}=\frac{\lambda}{2} \left(x^{(N)}\right)^2-\frac{(-)^{N+1}}{2}\lambda \beta^{2N} x^2\;,
\end{equation}
with $N\in\mathbb{N}_0$ and $\lambda$, $\beta\in\mathbb{R}^+$ can be used to illustrate some 
features of HD dynamics. 
It reduces to the standard harmonic oscillator for $N=1$, in which case $\lambda$ is a 
mass-parameter. The case $N=2$ has already been used as a toy model in \cite{Smilga05}.

The equation of motion (\ref{eom}) reads 

\begin{equation}
x^{(2N)}+\beta^{2N} x=0\,,
\end{equation}
and its classical solution is given by 
\begin{equation}
x(t)=\sum^{2N-1}_{j=0} A_j \, {\rm e}^{\beta_j t}\quad {\rm  with}\quad
\beta_j=\beta\, {\rm e}^{i\, \theta_j}\quad {\rm  and} \quad \theta_j=\frac{\pi}{2N}+\frac{j\pi}{N}\;.
\end{equation}
It can be observed from $x(t)$ that the allowed trajectories go beyond 
a standard periodic motion since, up to an appropriate choice of $A_j\,$, 
there may appear: 
\begin{itemize}
\item unbounded when ${\rm Re}\, \beta_j >0\,$. This occurs for $j$ such that 
$0<2j+1<N$ and $3N<2j+1\leqslant 4N-1\,$;
\item damped when ${\rm Re}\,  \beta_j<0\,$. This occurs for $j$ such that
$N<2j+1< 3N\,$;
\item periodic with period $2\pi/\beta$ if ${\rm Re}\, \beta_j=0$: This 
can happen for $N$ odd and the two values $j=(N-1)/2$ and $j=(3N-1)/2\,$.
\end{itemize}
The standard case $N=1$ is the only value for which all the possible trajectories are periodic. When $N>1$, damping or ``blowing-up" phenomena occur at time scales of order $\beta\cos(\pi/2N)\,$.

\subsection{General case: Pais-Uhlenbeck oscillator}

Lagrangian (\ref{ltoy}) 
always allows for unbounded classical motions. It can be showed that adding intermediate derivatives $x^{(0<i<N)}$ may reduce and even suppress such instabilities \footnote{Adding degrees of freedom coupled to $x$ is another way of addressing the problem, see. \textit{e.g.} \cite{moto}}. Lagrangian (\ref{ltoy}) may be generalised as follows
\begin{equation}\label{lgen}
L(x^{(0)},x^{(1)},\ldots , x^{(N)})
=\tfrac{1}{2}\,\sum^N_{j=0}a_j (x^{(j)}){}^{2}\;,\quad 
a_j\in\mathbb{R}\;,
\end{equation}
and the corresponding equation of motion reads
 \begin{equation}
\sum^N_{j=0} (-)^j a_j\, x^{(2j)}=0 \;.
\end{equation}
The characteristic polynomial of the above differential equation is
\begin{equation}\label{pcar}
p(\omega)=\sum^N_{j=0} (-)^j a_j\, \omega^{2j}\,.
\end{equation}
If the signs of $a_i$ are alternating, \textit{i.e.} $a_j=(-)^{j+1} b_j$ with $b_j\in\mathbb{R}^+_0\,$, then $p(\omega)$ can be factorised under the form $p(\omega)\sim \Pi^N_{j=1}\left(\omega^2-(i\omega_j)^2\right)$ 
with $\omega_j\in\mathbb{R}_0\,$ and all the trajectories $x(t)$ will be bounded. The further choice  $a_0<0$ is such that a standard potential energy is recovered 
for $N=1\,$. 

Therefore, replacing $(x^{(j)})^2$ by $(-)^jx\, x^{(2j)}$ up to total derivatives, 
Lagrangian (\ref{lgen}) can be rewritten as
\begin{equation}\label{lgen2}
L(x^{(0)},x^{(1)},\ldots , x^{(N)})=-\tfrac{1}{2}\, x \sum^N_{j=0}b_j x^{(2j)} \;.
\end{equation}
Lagrangian (\ref{lgen2}) is nothing but the P-U oscillator~\cite{Pais:1950za}, 
originally written under the equivalent form
\begin{equation}\label{L1}
L=-\tfrac{1}{2}\,x \,F_N\left(\frac{d}{dt}\right) x\quad {\rm where }\quad F_N(D)=\prod\limits^{N-1}_{i=0}\left(1+\frac{D^2}{\omega_i^2}\right)
\end{equation}
where the frequencies $\omega_i$ are assumed to be real and distinct. 
It has been shown in \cite{Pais:1950za} that equal or imaginary frequencies 
lead to unbounded trajectories so these cases will not be considered 
in the following. 
 
The solution of the equation of motion related to (\ref{L1}), $F_N\left(\frac{d}{dt}\right)x=0\,$, reads $x(t)=\sum^N_{k=1} A_k\, \sin(\omega_k\, t+\varphi_k)$ with  $A_k, \varphi_k\in\mathbb{R}\,$. All classical trajectories are therefore bounded. Note that $x(t)$ may describe the motion of a given mass in an $N-$body coupled harmonic oscillator whose normal modes have frequencies $\omega_i$: The formal analogy between the P-U oscillator and the dynamics of a $N-$body spring-mass system has been explored in the $N=2$ case in Ref. \cite{meca}.

An equivalent writing of (\ref{L1}) makes use of the oscillator variables 
\begin{equation}\label{Qdef}
     Q_i :=\prod\limits^{N-1}_{0=j\neq i} \left(1+\frac{1}{\omega_j^2}\frac{d^2}{dt^2}\right)x\
 \end{equation}
and shows that (\ref{L1}) can formally be written as a 
Lagrangian describing $N$ decoupled harmonic oscillators:
\begin{equation}\label{LPU}
    L=-\frac{1}{2}\sum^{N-1}_{i=0} \eta_i\, Q_i\left(\omega^2_i+\frac{d^2}{dt^2}\right)Q_i\;,
\end{equation}
where it can be deduced from \cite{Pais:1950za}, see \cite{Ketov:2011re}, that
\begin{equation}
\label{expliciteta}
   \eta_i=\frac{1}{\omega_i^2}\frac{\sum\limits_{\substack{0\leq m_0<\dots<m_{N-2}\leq N-1\\ m_j\neq i }}\omega^2_{m_0}\dots \omega^2_{m_{N-2}}}{\prod\limits^{N-1}_{\substack{m=0 \\ m\neq i}}(\omega^2_m-\omega^2_i)}\;~.
\end{equation}
The signs of $\eta_i$ are alternating, which is a typical signature of HD dynamics, 
eventually leading to a total energy whose sign is undefined. The equation of motion 
for the $Q_i$'s reads
\begin{align}
    \ddot{Q}_i + \omega_i^2 Q_i = 0\;, \qquad i=0,\ldots, N-1\;.
\end{align}

Generalization of the P-U oscillator in more than one spatial dimension 
is straightforward and will be left for future works. 
New integrals of motion such as HD angular momenta are naturally expected: 
we refer the interested reader to \cite{Riahi} for explicit definitions.

%*********************************************
\section{Hamiltonian formalism}\label{sec:ham}
%*********************************************

The Ostrograski construction that we have reviewed in Sec. \ref{sec:gene} 
canonically leads to Hamiltonians that are not separable in the variables $p_i$ (\ref{momenta}) and 
\begin{align}
    q^{i} &:= x^{(i)}\;,\quad \forall ~i \in\{ 0, \ldots , N-1\}\; .
\end{align}
In this section we show in some particular cases that a suitable change of canonical variables allows to recast the P-U Hamiltonian under a separable form. 

\subsection{Ostrogradski's approach}

As in the standard $N=1\,$ case, Hamilton's variational principle 
naturally leads to a symplectic structure and an Hamiltonian function $H\,$.
Provided that the regularity condition \eqref{regularity} holds 
one can invert the 
last relation \eqref{momenta} defining the momentum $p_{N-1}$ in order to 
express $x^{(N)}$ in terms of $p_{N-1}$ and the $x^{(i)}$'s, 
$i\in\{0,\ldots,N-1\}\,$: 
\begin{align}
\frac{\partial^2 L}{\partial x^{(N)}\partial x^{(N)}} \neq 0 \qquad 
\Rightarrow\qquad 
x^{(N)}={\cal V}(x^{(0)},\ldots, x^{(N-1)},p_{N-1})\;.    
\end{align}
Details about singular HD Lagrangians can be found in 
\cite{Nesterenko:1987jt,DuninBarkowski:2008uj}. 
The Hamiltonian function $H$ is defined as
\begin{align}
    H(\{q^{i},p_{i}\}_{i=0,\ldots, N-1}) := 
    \sum_{j=0}^{N-2} p_{i} q^{i+1} + p_{N-1}{\cal V} 
    - L(q^{0},q^{1},\ldots q^{N-1}, {\cal V})\;.
    \label{Hamilton}
\end{align}
The symplectic structure, as already apparent from the structure of the boundary terms 
in the variation $\delta S$ in the equation above \eqref{momenta},
is given by the two-form
$\mathbf{\Omega} = \sum_{i=0}^{N-1} dp_{i}\wedge dq^{i}\,$.
In particular, the Poisson-Ostrogradski bracket between any two functions 
in the phase space $T^*\mathbb{Q}$ locally coordinatized by the $2N$ variables
$(q^{i},p_{i})_{i=0,\ldots,N-1}\,$ is given by 
\begin{align}
    \{f,g\} = \sum_{i=0}^{N-1} \left( \frac{\partial f}{\partial q^{i}}
    \frac{\partial g}{\partial p_{i}} - 
    \frac{\partial f}{\partial p_{i}} \frac{\partial g}{\partial q^{i}}
    \right) \;.
    \label{PoissonBra}
\end{align}
The symplectic structure $\mathbf{\Omega}$ amounts to writing 
$\{q^{i},p_{j}\} = \delta^i_j\;,~\{q^{i},q^{j}\} \,=\, 0 \,=\, \{p_{i},p_{j}\}\;, ~
    \forall ~ i,j\in\{0,\ldots,N-1 \}\;$,
and the field equations are given as usual by 
$\dot{q}^{i} = \{q^{i},H\}\equiv \frac{\partial H}{\partial p_{i}}\;,~  \dot{p}_{i} = \{p_{i},H\} \equiv - \frac{\partial H}{\partial q^{i}}\;$.
The original equation of motion is reproduced from 
$\dot{p}_{0} = \{p_0,H\}\,$, while all the other equations above 
reproduce the relations \eqref{momenta} between 
the momenta- and the position-like variables. 

Starting from Lagrangian (\ref{lgen}) and applying Ostrogradski's procedure one gets 
\begin{equation}
p_i=\sum^{N-i-1}_{j=0}(-)^j a_{i+j+1} x^{(i+2j+1)}, \quad i=0,\dots,N-1    
\end{equation}
and the Hamiltonian reads
\begin{equation}\label{ham1}
H=    \sum_{j=0}^{N-2} p_{j} q^{j+1}+\frac{p^2_{N-1}}{2a_N}-\frac{1}{2}\sum^{N-1}_{j=0}a_j (q^j)^2.
\end{equation}
The equations of motion are explicitly given by
\begin{eqnarray}
    \dot q^i&=&q^{i+1},\quad i=0,\dots,N-2 \nonumber\\
    \dot q^{N-1}&=&\frac{p_{N-1}}{a_N} \\
    \dot p_0&=&a_0 q_0 \nonumber \\
    \dot p_j&=&-p_{j-1}+a_j q_j\,, \qquad j=1,\dots,N-1 \nonumber \;.
\end{eqnarray}
    
\subsection{Link with Pais-Uhlenbeck variables}

The equivalence of Ostrogradski and P-U formalisms at Hamiltonian level is not obvious
in the sense that the canonical transformation between the Ostrogradski 
and the P-U variables has not been explicitly given in the general case, 
as far as we could see.
It was shown for $N=2$ in \cite{pagani}. Here we show it for the $N=3$ case 
and leave the explicit form of the canonical transformation relating the 
Ostrogradski to the P-U variables for future works. 
Below we simply give the general expression for the generating function of the 
canonical transformation for arbitrary $N\,$, without solving the partial 
differential equations that explicitly relate the two sets of phase space
variables. 

%and then generalize our conclusions to arbitrary $N\,$.

\subsubsection{The case N=3}
A convenient parameterisation in this case is
\begin{equation}\label{LN3}
    L = -\frac{1}{2}\, x^2+\frac{A_1}{2\Omega^6}\dot x^2-\frac{A_2}{2\Omega^6} \ddot x^2 +\frac{1}{2\Omega^6} \dddot x^2\;,
\end{equation}
with 
\begin{equation}
A_1:=\sum_{i<j}(\omega_i\omega_j)^2,\quad 
A_2:=\sum_{i=0}^{2}\omega_i^2\;,\quad      
\Omega^6:=(\omega_0\omega_1\omega_2)^2\quad {\rm and} \quad \omega_2>\omega_1>\omega_0\;.
\end{equation}
Note that Lagrangian (\ref{LN3}) can be recast under the form (\ref{LPU}) 
provided that 
\begin{equation}
    \sum^2_{i=0}\omega^2_i\eta_i=1 \;,\quad \sum^2_{i=0}\omega^4_i\eta_i=0=\sum^2_{i=0}\omega^6_i\eta_i \;.\\
\end{equation}
The above constraints are fulfilled with (\ref{expliciteta}). The momenta associated with (\ref{LN3}) read 
\begin{align}
\label{pOstrogra}
    p_0 &= \frac{1}{\Omega^6}\left( A_1 \dot x+A_2 \dddot x+x^{(5)}\right)\;, \quad
    p_1 = -\frac{1}{\Omega^6}\left(A_2\ddot x+x^{(4)}\right) \;, \quad
    p_2 =\frac{\dddot x}{\Omega^6}
\end{align}
and the Hamiltonian is given by
\begin{equation}
    H=p_0 q^1+p_1 q^2+\frac{\Omega^6}{2}p_2^2+\frac{1}{2}(q^0)^2-\frac{A_1}{2\Omega^6}(q^1)^2+\frac{A_2}{2\Omega^6}(q^2)^2\;.
\end{equation}
One can readily derive a generating function for the canonical 
transformation relating the Ostrogradski variables $(q^0,q^1,q^2,p_0,p_1,p_2)$ 
to the P-U phase space variables $(Q_0,Q_1,Q_2,P_0,P_1,P_2)\,$.
It is given by 
\begin{align}
    F(q^0,p_1,q^2,P_0,P_1,P_2) &= \sum_{i=0}^2 Q_i(q^0,p_1,q^2) P_i 
    \,\equiv\, \sum_{i=0}^2 \tilde{Q}_i(q^0,p_1,q^2) \tilde{P}_i\;.
\end{align}
It leads to the following system of partial differential equations:
\begin{align}
    \frac{\partial F}{\partial P_0} & = Q^0(q^0,p_1,q^2)\;,\qquad 
    \frac{\partial F}{\partial P_1} =  Q^1(q^0,p_1,q^2)\;,\qquad 
    \frac{\partial F}{\partial P_1} =  Q^1(q^0,p_1,q^2)\;,
    \\
    \frac{\partial F}{\partial q^0} & = p_0(q^0,p_1,q^2)\;,\qquad 
    \frac{\partial F}{\partial p_1} = - q^1(q^0,p_1,q^2)\;,\qquad
    \frac{\partial F}{\partial q^2} = p_2(q^0,p_1,q^2)\;.
\end{align}
The first three equations are trivial, whereas the last three can be obtained 
from the relations $\sum_{i=0}^2 \eta_i\,\omega_i^{2m} = \delta^m_1\;,$ 
$m=1,2,3\;$, 
as well as 
$\sum_{i=0}^2 \eta_i \omega_i^8 = \omega_0^2\omega_1^2\omega_2^2\;$
and $\sum_{i=0}^2 \eta_i = \frac{A_1}{\Omega^6}\,$.
In particular, it yields
\begin{align}
    p_0 &= P_0+P_1+P_2  \;,\qquad 
    p_2 = -\frac{1}{\Omega^6}\,\sum_{i=0}^2 \omega_i^4 P_i  \;,\qquad 
    q^1 = \sum_{i=0}^2 \omega_i^2 P_i\;.  
\end{align}
Recalling that the $\eta_i$ are of alternating sign, 
the Ostrogradski Hamiltonian eventually reads 
\begin{equation}
    H=\tfrac{1}{2}\sum^2_{i=0}(-)^i\left(\frac{P^2_i}{\vert\eta_i\vert}\,
    +\vert\eta_i\vert\,\omega^2_i\, Q_i^2\right)\;.
\end{equation}
Let us observe that, from the expressions \eqref{pOstrogra} for the Ostrogradski 
momenta and the definition \eqref{Qdef} of the P-U oscillator variables $Q_i\,$, 
$i=0,1,2\,$, one has the relation $P_i = \eta_i\,\dot{Q}_i\;$, $\;i=0,1,2\;$.
Finally, a last canonical transformation given by the mere rescaling  
$\tilde{Q}_i := \sqrt{|\eta_i|}\,Q_i\;$, $\;\tilde{P}_i 
    := \frac{P_i}{\sqrt{|\eta_i|}}\;$, 
gives the following P-U Hamiltonian:
\begin{equation}
    H=\tfrac{1}{2}\sum^2_{i=0}(-)^i\left(\tilde{P}^2_i + \omega^2_i \tilde{Q}_i^2\right)\;.
\end{equation}

\subsubsection{Arbitrary N}
The $N=2$ and $N=3$ cases show that Ostrogradski's procedure leads to P-U 
Hamiltonian and variables by successive canonical transformations. Moreover Lagrangian (\ref{lgen}), once expressed as (\ref{LPU}), leads to the separable Hamiltonian
\begin{equation}\label{hamPU}
    H=\tfrac{1}{2}\sum^{N-1}_{i=0}(-)^i\left(\frac{P^2_i}{\vert\eta_i\vert}\,
    +\vert\eta_i\vert\,\omega^2_i\, Q_i^2\right)=:\sum^{N-1}_{i=0} (-)^iE_i \;
\end{equation}
for any $N$, where $E_i\,$, $i=0,1,2\,$ are positive definite quantities and where 
\begin{align}
Q_i &:=\prod\limits_{j\neq i} 
\left(1+\frac{1}{\omega_j^2}\frac{d^2}{dt^2}\right)x\;,
\label{PUQ}\\
P_i &:=\eta_i\,\dot Q_i \equiv (-)^i\vert\eta_i\vert\, \dot Q_i\; ,
\end{align}
and where the P-U coefficients $\eta_i$ can be found in \cite{Ketov:2011re} 
and in \eqref{expliciteta}. Note that the frequencies $\omega_k$ are the roots of the polynomial equation $p(\omega_k)=0$, see (\ref{pcar}).

%*******************************************
\section{Adiabatic invariants and Nekhoroshev estimates}\label{sec:ac-an}
%*******************************************

\subsection{Action variables}
The P-U Hamiltonian (\ref{hamPU}) is separable and admits elliptic 
trajectories in the planes $(Q_i,P_i)\,$, $i\in\{0,1,\ldots,N-1\}\,$, 
these fixed-$E_i$ cycles being denoted as $\Gamma_i\,$. 
Hence a set of $N$ action variables can be defined:
\begin{equation}\label{Idef}
I_j=\frac{(-)^j}{2\pi} \oint_{\Gamma_j} P_j\, dQ_j= \frac{\vert\eta_j \vert}{2\pi}  \oint_{\Gamma_j} \dot Q_j^2\, dt\;, \quad j=0,\dots,N-1\; .
\end{equation}
The $(-)^j$ factor is such introduced in such a way that the action variables 
$\{I_i\}_{i=0,\ldots, N-1}$ are all positive. 
It can be checked that the Hamiltonian (\ref{hamPU}) reads
\begin{equation}\label{PUH}
H=\sum^{N-1}_{j=0} (-)^j I_j \omega_j , \quad {\rm with}\quad   I_j=\frac{E_j}{\omega_j}\;,
\end{equation}
and that the relations
\begin{equation}\label{Idef2}
I_i=\frac{\partial E_i }{\partial \omega_i}\;,\quad i=0,\ldots, N-1 
\end{equation}
holds as well. 
The action variable $I_0$ reduces to the average kinetic energy 
for $N=1\,$, as expected. 

The action variables can be expressed in terms of the classical trajectory $x(t)$ through the definition (\ref{PUQ}). When $N=2$ for example, 
\begin{eqnarray}
    I_0&\sim &\oint_{\Gamma_0} \left[\dot x^2+\frac{2}{\omega^2_1}\dot x\, \dddot x+\frac{1}{\omega^4_1}\dddot x^2 \right] dt\;,\\
   I_1&\sim &\oint_{\Gamma_1} \left[\dot x^2+\frac{2}{\omega^2_0}\dot x\, \dddot x+\frac{1}{\omega^4_0}\dddot x^2 \right] dt\;.\nonumber
\end{eqnarray} 
The cross-product $\dot x\, \dddot x$ cannot be expressed in terms 
of $\ddot x^2$ with the use of a total derivative since $x(t)$ is not 
a priori periodic with frequency $\omega_0$ or $\omega_1\,$ unless 
$\omega_1/\omega_0=n/m\in\mathbb{Q}\,$. 
In the latter case, after a time 
$T = m 2\pi/\omega_0=n 2\pi /\omega_1\,$, the action variables can be recast 
under the form $\oint_{\Gamma_0} \left[\dot x^2+c \ddot x^2+d \dddot x^2 \right] dt$ with $c,d$ real coefficients. 
At the same time, the commensurability condition on $\omega_1/\omega_0$ implies 
instability of the $N=2$ dynamics under small perturbations so it is not relevant for 
the present study \cite{pagani}. 

The $I_j$ are constant of motion provided that $H$ does not explicitly depend on time. It is nevertheless possible that some external parameter is time-dependent: we set  $\omega_i=\omega_i(t)$ in (\ref{hamPU}). Under the assumption that the $\omega_i(t)$ vary slowly enough with respect to the typical duration of a cycle -- and despite the fact that no rigorous definition of ``enough" can be given \cite{HA} --,  the quantities $I_j$ given by (\ref{Idef}) are adiabatic invariants. It can be deduced from \cite{LL} that their small rate of change is given by:
\begin{eqnarray}\label{ac-an}
\dot I_k&=&-\frac{\dot \omega_k}{\omega_k}I_k \cos 2\phi_k\\
\dot\phi_k&=&(-)^k\omega_k+\frac{\dot\omega_k}{2\omega_k}\sin 2\phi_k, \nonumber
\end{eqnarray}
where $\phi_k$ are the angle variables conjugated to $I_k$: $\{\phi_j,I_k \}=\delta_{jk}$. The interested reader may find general computations related to time-varying harmonic oscillators in \cite{wells}. 

As an illustration of the above relations, suppose that the time-dependent parameter 
is a small perturbation of the frequencies:
\begin{equation}\label{pert}
    \omega_i(t)=\varpi_{i} ( 1 + \epsilon \,g_i(t) )\;,
    \quad {\rm with} \quad \vert  \epsilon \vert \ll 1\; ,
\end{equation}
the only dependence on time being contained in the real functions $g_i\,$. 
At the lowest order in $\epsilon$, Eqs. (\ref{ac-an}) become
\begin{eqnarray}\label{I2}
&\dot I_k = -\,\epsilon\,\dot{g_k}\,I_k\cos{2\varphi_k}\;,
\quad 
\dot{\phi}_k = (-)^k\varpi_k(1+\epsilon\, g_k) +
\frac{\epsilon}{2}\,\dot{g}_k\,\sin{2\varphi_k}\;,&
 \\
& \varphi_k := (-)^k\varpi_k t +\alpha_k\;& .
\end{eqnarray}
For $g_i(t)$ arbitrary and $\bar I_k$ the
values of the action variables for the unperturbed system, 
the simple shape 
\begin{equation}
    I_k=\bar I_k ( 1- \epsilon\,g_k )
\end{equation}
solves Eq. (\ref{I2}) at the first order in $\epsilon$ when $\varphi_k=n\, \pi\,$, $n\in\mathbb{Z}\,$. It may therefore be used to estimate the trend of the modifications induced by $g_i$ on the action variables.

In the special case $g_i(t) = \bar g \,\sin{(\Omega_i t +\beta_i)}\,$, 
the solution of (\ref{I2}) up to first order in $\epsilon$ is given by   
\begin{align}
    I_k &= {\bar I}_k \left[1-\frac{\epsilon}{2}\;{\bar g} \,
    \Big(\frac{\Omega_k}{2\omega_k^+}\,\sin{2(\omega_k^+ t +\alpha_k^+)}
    + \frac{\Omega_k}{2\omega_k^-}\,\sin{2(\omega_k^- t +\alpha_k^-)}\Big)
    \right]\;,\\
    \phi_k &= \alpha_k + (-)^k\varpi_k t + (-)^{k+1}\,\epsilon\,\frac{\varpi_k}{\Omega_k}\;{\bar g}\,
    \cos{(\Omega_k t +\beta_k)}\\
    & \qquad + (-)^{k+1}\,\frac{\epsilon}{4}\;{\bar g}\,
    \Big(\frac{\Omega_k}{2\omega_k^+}\,\cos{2(\omega_k^+ t + \alpha_k^+)}
    + \frac{\Omega_k}{2\omega_k^-}\,\cos{2(\omega_k^- t + \alpha_k^-)}
    \Big)\;, \nonumber \\
    & \omega_k^\pm := \varpi_k \pm \tfrac{(-)^k}{2}\,\Omega_k\;,\quad 
    \alpha_k^\pm := (-)^k(\alpha_k \pm \frac{\beta_k}{2}\,)\;.
\end{align}
In the case where the perturbations  $g_k(t)$ 
are slow, as is prescribed by the
theory of adiabatic invariants, one has that the frequencies $\Omega_k$ are 
small compared to the characteristic frequencies $\varpi_k$ of the system and 
one finds that the variation of the action variables starts at 
second order, while the variation of the angle variables starts earlier. 

%\begin{eqnarray}\label{model}
%I_k &=& \bar I_k\left[ 1 - 
%\frac{\epsilon}{2\varpi_k}\,\bar g\,
%\Big(\frac{\Omega_k}{\Omega_k^+}\sin{(\Omega_k^+ t+\beta_k+2\alpha_k)}
%+
%\frac{\Omega_k}{\Omega_k^-}\sin{(\Omega_k^- t+\beta_k-2\alpha_k)}
%\Big)\right]\; ,
%\\
%\Omega_k^{\pm} &:=& \Omega_k \pm 2(-)^k\varpi_k\;, \nonumber
%\\
%\phi_k &=& \alpha_k+(-)^k\varpi_k t
%- (-)^k \epsilon {\bar g} 
%\frac{\varpi_k}{\Omega_k}\,\cos{(\Omega_k t+\beta_k)} \nonumber \\
%&&-\frac{\epsilon}{4}\,{\bar g} 
%\Big(
%\frac{\Omega_k}{\Omega^+_k}\cos{(\Omega^+_k t+\beta_k+2\alpha_k)}- %\frac{\Omega_k}{\Omega^-_k}\cos{(\Omega^-_k t+\beta_k-2\alpha_k)}
%\Big)
%\end{eqnarray}
%

\subsection{Nekhoroshev estimates}

The P-U Hamiltonian (\ref{PUH}) with the time-dependent perturbation (\ref{pert}) can be formally written under the form 
\begin{equation}\label{hIphi}
H=h(\vec{I})+\epsilon f(\vec{ I},\vec{ \phi}),
\end{equation}
where 
\begin{equation}
h(\vec{ I})=\vec{ \varpi}\cdot \vec{  I},\quad f(\vec{  I},\vec{ \phi})=\vec{  I}\cdot \vec{  g}(\vec{  \phi}),
\end{equation}
and where the vectors $(\vec{  I})_k= I_k$, $(\vec{  \phi})_k=\phi_k$, $(\vec{ g})_k=g_k$ and $(\vec{  \varpi})_k=(-)^k\varpi_k$ have been introduced. %It is assumed that the equations of motion (\ref{ac-an}) may be used to replace $g_k(t)$ by $g_k(\phi_k)$. 

Nekhoroshev's theorem \cite{nekho} states that if the nearly integrable Hamiltonian (\ref{hIphi}) is analytic and the unperturbed part $h(\vec{ I})$ is steep (or convex, or quasiconvex) on some 
domain, then there is a threshold $\epsilon_0>0\,$ and positive constants $R$, $T$,
$a$ and $b$ such that whenever $|\epsilon|<\epsilon_0\,$, for all initial actions variables
$\vec I(0)$ in the domain (and far enough from the boundary), one has
\begin{align}
    \max\{\vert I_k(t)-I_k(0)\vert \}< R\,\epsilon^b\qquad {\rm for~times}\qquad
    |t|<T \exp{\varepsilon^{-a}}\;.
\end{align}
This result has been particularized to several explicit examples, among which the harmonic oscillator in Refs. \cite{benett,valdi}. The fact that some components of $\vec{  \varpi}$ are negative is allowed by the formalism of the latter references. From the study \cite{benett} in particular, it can be deduced that unstable behaviours in the P-U oscillator appears at exponentially large time in $\epsilon$, \textit{i.e.} the dynamics is classically stable. Since $\epsilon_0\sim N^{-2N}$ \cite{benett}, the more the dynamics contains HD, the more the perturbation must be small to preserve stability. Moreover, there must exist two real positive constants $\sigma$, $\tau$ such that  $\vert \vec\varpi\cdot \vec n\vert \geq \sigma  \left[ \sum^{N-1}_{j=0}\vert n_j \vert\right]^{-\tau} $ for all $\vec n\in \mathbb{Z}^N_0$ otherwise $\epsilon_0$ becomes arbitrarily large and the system is unstable. 
In other words the frequencies must define a non resonant harmonic oscillator. Such an instability can only occur for $N>1$ one-dimensional dynamics; it is trivially avoided when $N=1$ because no energy transfer between the different components $E_i$ are \textit{de facto} absent in this case.

%*******************************************
\section{Concluding comments}\label{sec:conclu}
%*******************************************

Higher-derivative action principles generally lead to unstable classical dynamics. 
However, all the classical trajectories allowed by the Pais-Uhlenbeck oscillator (\ref{L1}) 
with distinct and nonresonant frequencies are bounded: It is an explicit realisation of a 
stable classical theory with higher-derivatives. Therefore the problem can be formulated in 
action-angle variables formalism, allowing a computation of adiabatic invariants and a proof 
of the classical stability based on Nekhoroshev estimates. Although the Pais-Uhlenbeck 
oscillator has been widely studied as a prototypal higher-derivative physical theory, it is 
the first time, to our knowledge, that such results are obtained. 
Emphasis has been put on harmonic potentials in the present study. 
Other types of potentials or higher-derivative terms may also lead to stable classical 
dynamics. For example it is shown in Ref. \cite{pagani} that a $N=2$ Pais-Uhlenbeck 
oscillator with cubic and quartic potential terms is stable too, except for very special 
values of the parameters. 
The stable models of \cite{pagani} should give, after generalisation to field theory 
Lagrangians, further examples (compared to the one reviewed in \cite{Smilga:2017arl}) 
of well-behaved dynamical systems with infinite number of degrees of freedom. 

One can recover a Lagrangian function from a Hamiltonian function
provided that the regularity condition
det$\,\frac{\partial^2 H}{\partial p_{N-1}\partial p_{N-1}}\neq 0\,$ is satisfied
--- which is the case for a free system that is regular when 
the added interactions in the perturbed Hamiltonian are algebraic functions in the 
P-U variables $Q$'s, as we assumed implicitly here. 
This is clear for $N$ odd since the latter variables 
do not imply Ostrogradski's last momentum $p_{N-1}\,$. 
In the even $N$ cases where the $Q$ variables involve $p_{N-1}\,$, the 
added perturbations cannot ruin the regularity property of the 
unperturbed P-U Hamiltonian as long as the perturbation is 
of polynomiality degree in $Q$ higher than two, as we assume here. 
However, 
it is notoriously difficult (see e.g. Section 8 of \cite{Smilga:2017arl}) to 
find the analytical expression  of the perturbed Lagrangian  corresponding 
to a given Hamiltonian perturbation. It it usually nonlocal, even for simple
(e.g. quartic) Hamiltonian perturbation \cite{Smilga:2017arl}, 
which suggests considering as a starting 
point the $N\rightarrow \infty$ limit studied by Pais and Uhlenbeck, as it would 
encompass all the possible Hamiltonian perturbations.

It is worth making comments about quantization. In a first approximation a Bohr-Sommerfeld 
quantization rule can be applied since action variables exist. The fact that the energy 
spectrum is unbounded both from below and above in higher-derivative theories does not a 
priori forbids well-behaved quantum dynamics. 
In fact, a quantization technique was proposed in \cite{Kaparulin:2014vpa} that keeps the 
higher-derivative dynamics stable at the quantum level. 
In fact, we propose that the positive-definite quantities suggested in  
\cite{Kaparulin:2014vpa} and that are responsible for a stable dynamics at the quantum 
level are nothing but the action variables. 
Indeed, even for a perturbed classical motion, as long as the 
trajectories are bounded, the action variables are 
positive-definite quantities and conserved to the approximation 
given.
More recently, it was conjectured in \cite{Smilga:2017arl} that indeed, 
when all the classical trajectories of a 
given higher-derivative model are bounded, its quantum dynamics only contain 
so-called benign ghosts, \textit{i.e.} negative-energy quantum states with a normalisable 
wave function and preserved unitarity of the evolution. 
The present work aimed at clarifying the conditions for higher-derivative models 
to exhibit bounded classical dynamics; hence it is a step toward the identification 
of quantum models with unitary quantum dynamics, to which the Pais-Uhlenbeck oscillator 
belongs. Further issues about quantization of higher-derivative Lagrangians are 
discussed for example in \cite{gitman}. 

Finally it has to be noticed that the necessary conditions for adiabatic invariants and Nekhoroshev estimates to be computed are the separability of the higher-derivative Hamiltonian and the existence of bounded classical trajectories. Both conditions are met in the Pais-Uhlenbeck oscillator case after appropriate choice of canonical variables, but we believe that other classes of higher-derivative systems may be studied by resorting to the methods we have presented.

%\appendix
%\section{Conventions}

%\bibliographystyle{utphys.bst}
%\bibliography{biblio}

\begin{thebibliography}{9}
%
\bibitem{Pais:1950za} 
  A.~Pais and G.~E.~Uhlenbeck, \textit{On Field theories with nonlocalized action},
  Phys.\ Rev.\  {\bf 79}, 145 (1950).
%%CITATION = doi:10.1103/PhysRev.79.145;%%
%
\bibitem{Nesterenko:2006tu} 
  V.~V.~Nesterenko,
  \textit{On the instability of classical dynamics in theories with higher derivatives},
  Phys.\ Rev.\ D {\bf 75}, 087703 (2007)
%  doi:10.1103/PhysRevD.75.087703
  [hep-th/0612265].
%%CITATION = doi:10.1103/PhysRevD.75.087703;%%
%  
\bibitem{Smilga:2008pr} 
  A.~V.~Smilga, \textit{Comments on the dynamics of the Pais-Uhlenbeck oscillator},
  SIGMA {\bf 5}, 017 (2009)
  [arXiv:0808.0139].
%%CITATION = doi:10.3842/Sigma.2009.017;%%  
%
\bibitem{Joung:2012sa}
  E.~Joung and K.~Mkrtchyan,
  \textit{Higher-derivative massive actions from dimensional reduction},
  JHEP {\bf 1302} (2013) 134
%  doi:10.1007/JHEP02(2013)134
  [arXiv:1212.5919].
%%CITATION = doi:10.1007/JHEP02(2013)134;%%
%
\bibitem{Pavsic:2013mja} 
  M.~Pavšič, \textit{Pais-Uhlenbeck Oscillator with a Benign Friction Force},
  Phys.\ Rev.\ D {\bf 87}, no. 10, 107502 (2013)
  [arXiv:1304.1325].
%%CITATION = doi:10.1103/PhysRevD.87.107502;%% 
%
\bibitem{Kaparulin:2014vpa} 
  D.~S.~Kaparulin, S.~L.~Lyakhovich and A.~A.~Sharapov,
  \textit{Classical and quantum stability of higher-derivative dynamics},
  Eur.\ Phys.\ J.\ C {\bf 74}, no. 10, 3072 (2014)
  [arXiv:1407.8481].
%%CITATION = doi:10.1140/epjc/s10052-014-3072-3;%%  
%
\bibitem{Woodard:2015zca} 
  R.~P.~Woodard, \textit{Ostrogradsky's theorem on Hamiltonian instability}, Scholarpedia {\bf 10}, no. 8, 32243 (2015) [arXiv:1506.02210].
%%CITATION = doi:10.4249/scholarpedia.32243;%%  
%  
\bibitem{Masterov:2016jft} 
  I.~Masterov,
\textit{The odd-order Pais–Uhlenbeck oscillator},
  Nucl.\ Phys.\ B {\bf 907}, 495 (2016)
  [arXiv:1603.07727].
%%CITATION = doi:10.1016/j.nuclphysb.2016.04.025;%%  
%
\bibitem{Avendano-Camacho:2017ykt} 
  M.~Avendaño-Camacho, J.~A.~Vallejo and Y.~Vorobiev,
\textit{A perturbation theory approach to the stability of the Pais-Uhlenbeck oscillator},
  J.\ Math.\ Phys.\  {\bf 58}, no. 9, 093501 (2017)
  [arXiv:1703.08929].
%%CITATION = doi:10.1063/1.5000382;%%  
%
\bibitem{Smilga:2017arl} 
  A.~Smilga,
  \textit{Classical and quantum dynamics of higher-derivative systems},
  Int.\ J.\ Mod.\ Phys.\ A {\bf 32}, no. 33, 1730025 (2017)
  [arXiv:1710.11538].
%%CITATION = doi:10.1142/S0217751X17300253;%%  
%
\bibitem{Nicolis:2018lkg} 
  S.~Nicolis, \textit{Higher time derivatives in the microcanonical ensemble describe dynamics of flux-coupled classical and quantum oscillators},
  arXiv:1805.07934 [hep-th].
  %%CITATION = ARXIV:1805.07934;%%
\bibitem{Horvathy:2002vt} 
  P.~A.~Horvathy and M.~S.~Plyushchay,
\textit{Non-relativistic anyons, exotic Galilean symmetry and noncommutative plane},
  JHEP {\bf 0206}, 033 (2002)
  doi:10.1088/1126-6708/2002/06/033
  [hep-th/0201228].
  %%CITATION = doi:10.1088/1126-6708/2002/06/033;%%
  \bibitem{Plyushchay:2006pw} 
  M.~S.~Plyushchay,
\textit{Majorana equation and exotics: Higher derivative models, anyons and noncommutative geometry},
  Electron.\ J.\ Theor.\ Phys.\  {\bf 3}, no. 10, 17 (2006)
  [math-ph/0604022].
  %%CITATION = MATH-PH/0604022;%%
  
\bibitem{Stelle:1976gc}
K.~S.~Stelle, \textit{Renormalization of Higher Derivative Quantum Gravity},
Phys.\ Rev.\ D {\bf 16}, 953 (1977);\textit{Classical Gravity with Higher Derivatives}, 
Gen.\ Rel.\ Grav.\  {\bf 9}, 353 (1978).
%%CITATION = doi:10.1103/PhysRevD.16.953;%%
%
\bibitem{Nitta:2018yzb} 
M.~Nitta and R.~Yokokura,
\textit{Higher derivative three-form gauge theories and their supersymmetric extension},
JHEP {\bf 1810} (2018) 146 
%  doi:10.1007/JHEP10(2018)146
[arXiv:1809.03957].
%%CITATION = doi:10.1007/JHEP10(2018)146;%%
%
\bibitem{Giribet:2018hck} 
G.~Giribet, O.~Miskovic, R.~Olea and D.~Rivera-Betancour, 
\textit{Energy in Higher-Derivative Gravity via Topological Regularization},
Phys.\ Rev.\ D {\bf 98}, 044046 (2018) [arXiv:1806.11075].
%%CITATION = doi:10.1103/PhysRevD.98.044046;%%
%
\bibitem{Tomboulis:1980bs} E. Tomboulis, 
\textit{Renormalizability and asymptotic freedom in quantum gravity}, Physics Letters B, 
\textbf{97}, 77--80 (1980). 
%
\bibitem{Schmidt:1994iz}
  H.~J.~Schmidt,
  \textit{Stability and Hamiltonian formulation of higher derivative theories}, 
  Phys.\ Rev.\ D {\bf 49} (1994) 6354
   Erratum: [Phys.\ Rev.\ D {\bf 54} (1996) 7906]
 % doi:10.1103/PhysRevD.49.6354, 10.1103/PhysRevD.54.7906
  [gr-qc/9404038].
  %%CITATION = doi:10.1103/PhysRevD.49.6354, 10.1103/PhysRevD.54.7906;%%
%
%\cite{Hawking:2001yt}
\bibitem{Hawking:2001yt}
  S.~W.~Hawking and T.~Hertog,
  \text{Living with ghosts},
  Phys.\ Rev.\ D {\bf 65} (2002) 103515
%  doi:10.1103/PhysRevD.65.103515
  [hep-th/0107088].
%%CITATION = doi:10.1103/PhysRevD.65.103515;%%
%
\bibitem{Ferrara:2018wqd} 
S.~Ferrara, A.~Kehagias and D.~Lüst, \textit{Aspects of Weyl Supergravity}, 
  JHEP {\bf 1808} (2018) 197
%  doi:10.1007/JHEP08(2018)197
  [arXiv:1806.10016].
%%CITATION = doi:10.1007/JHEP08(2018)197;%%
%
  \bibitem{Castellanos:2018dub} 
  A.~R.~R.~Castellanos, F.~Sobreira, I.~L.~Shapiro and A.~A.~Starobinsky, \textit{On higher derivative corrections to the $R+R^2$ inflationary model}, arXiv:1810.07787 [gr-qc].
%%CITATION = ARXIV:1810.07787;%%  
%
  \bibitem{motorC}
  W.L. Nelson, \textit{Physical Principles for Economics of Skilled Movements}, Biol. Cybern. \textbf{46}, 135--147 (1983); N. Hogan, \textit{An organizing principle for a class of voluntary movements}, The Journal of Neuroscience \textbf{4}, 2745--2754 (1984); S. Lebedev, W.H. Tsui and P. Van Gelder, \textit{Drawing Movements as an Outcome of the Principle of Least Action}, Journal of Mathematical Psychology \textbf{45}, 43--52 (2001); S. Hagler, \textit{On the Principled Description of Human Movements}, arXiv:1509.06981; D Huh and T.J. Sejnowski, \textit{Conservation law for self-paced movements}, PNAS \textbf{113}, 8831--8836 (2016).
%
  \bibitem{Ostro} 
  M.~Ostrogradsky, \textit{M\'{e}moires sur les \'{e}quations diff\'{e}rentielles, relatives au probl\`{e}me des isop\'{e}rim\`{e}tres},
  Mem.\ Acad.\ St.\ Petersbourg {\bf 6}, no. 4, 385 (1850).
%\cite{Ostrogradsky:1850fid}  
%
  \bibitem{benett} G. Benettin and G. Gallavotti, \textit{Stability of Motions near Resonances
in Quasi-Integrable Hamiltonian Systems}, Journal of Statistical Physics \textbf{44}, 293--338 (1986).
%
\bibitem{valdi} E. Valdinoci, \textit{Estimates for Non-Resonant  Normal Forms in
Hamiltonian Perturbation Theory}, Journal of  Statistical  Physics \textbf{101}, 4 (2000).
%
\bibitem{pagani} E. Pagani, G. Tecchiolli and S. Zerbini, \textit{On the Problem of Stability for Higher-Order Derivative Lagrangian Systems}, Lett. in Math. Phys. \textbf{14}, 311--319 (1987).
%\cite{Pagani:1987ue}
%%CITATION = doi:10.1007/BF00402140;%%
%
\bibitem{nekho} N.N. Nekhoroshev, \textit{Behavior of Hamiltonian systems close to integrable}, Functional Analysis and Its Applications. 5, 338--339 (1971); \textit{An exponential estimate of the time of stability of nearly-integrable Hamiltonian systems}, Russian Math. Surveys \textbf{32}, 1--65 (1977).
%
\bibitem{Smilga05} 
  A.~V.~Smilga, \textit{Ghost-free higher-derivative theory},
  Phys.\ Lett.\ B {\bf 632}, 433 (2006)
  [hep-th/0503213].
%%CITATION = doi:10.1016/j.physletb.2005.10.014;%%  
%\cite{Smilga:2005gb}
%
\bibitem{moto} 
  H.~Motohashi, K.~Noui, T.~Suyama, M.~Yamaguchi and D.~Langlois,
  \textit{Healthy degenerate theories with higher derivatives},
  JCAP {\bf 1607}, 033 (2016)
  [arXiv:1603.09355].
%%CITATION = doi:10.1088/1475-7516/2016/07/033;%%
%\cite{Motohashi:2016ftl}
%  
\bibitem{meca} N.G. Stephen, \textit{On the Ostrogradski instability for higher-order derivative theories and a pseudo-mechanical energy}, Journal of Sound and Vibration \textbf{310}, 729 (2008).
%
\bibitem{Ketov:2011re} 
  S.~V.~Ketov, G.~Michiaki and T.~Yumibayashi,
  \textit{Quantizing with a higher time derivative},
  Advances in Quantum Field Theory, InTech Publishers 2012, 49-73
  [arXiv:1110.1155 [hep-th]].
  %%CITATION = ARXIV:1110.1155;%%
%
   \bibitem{Riahi} M.~Borneas, \textit{Principle of action with higher derivatives}, Phys.\ Rev.\  {\bf 186}, 1299 (1969). F. Riahi, \textit{On Lagrangians with Higher Order Derivatives}, Am. J. Phys. {\bf 40}, 386 (1972); G.~C.~Constantelos, \textit{Integrals of motion for Lagrangians including higher-order derivatives}, Nuovo Cim.\ B {\bf 21}, 279 (1974).
%
   \bibitem{Nesterenko:1987jt} V.~V.~Nesterenko, \textit{The Singular Lagrangians With Higher Derivatives},
  J.\ Phys.\ A {\bf 22}, 1673 (1989).
  %%CITATION = doi:10.1088/0305-4470/22/10/021;%%
%  
  \bibitem{DuninBarkowski:2008uj}
    P.~Dunin-Barkowski and A.~Sleptsov,
  \textit{Geometric Hamiltonian Formalism for Reparametrization Invariant 
  Theories with Higher Derivatives}, Theor.\ Math.\ Phys.\  {\bf 158} (2009) 61
   [arXiv:0801.4293 [hep-th]].
  %%CITATION = doi:10.1007/s11232-009-0005-7;%%
% 
  \bibitem{HA} J.H. Hannay, \textit{Angle variable holonomy in adiabatic excursion of an integrable Hamiltonian}, J. Phys. A: Math. Gen. \textbf{18}, 221--230 (1985).
%
\bibitem{LL}   L. Landau and E. Lifchitz, \textit{Physique th\'{e}orique Tome 1 : M\'{e}canique} (4th Ed., MIR Moscou, 1988).
%
 \bibitem{wells} C.G. Wells and S.T.C. Siklos, \textit{The Adiabatic Invariance of the Action Variable in Classical Dynamics}, Eur. J. Phys. \textbf{28}, 105--112, 2007  [arXiv:physics/0610084v1]; Luis L. Sánchez-Soto and Jesús Zoido, \textit{Variations on the adiabatic invariance: The Lorentz pendulum},
American Journal of Physics \textbf{81}, 57 (2012);  R.M. Kulsrud, \textit{Adiabatic Invariant of the Harmonic Oscillator}, Phys. Rev. \textbf{106}, 205--207 (1957).  
%
\bibitem{gitman} 
D.\ M.\ Gitman, S.\ L.\ Lyakhovich and I.\ V.\ Tyutin, 
\textit{Hamilton formulation of a theory with high derivatives},
Soviet Physics Journal \textbf{26}, 730--734 (1983) ; %https://doi.org/10.1007/BF00898884
%\cite{Simon:1990ic}
  J.~Z.~Simon,
  \textit{Higher Derivative Lagrangians, Nonlocality, Problems and Solutions},
  Phys.\ Rev.\ D {\bf 41} (1990) 3720;
 % doi:10.1103/PhysRevD.41.3720
  %%CITATION = doi:10.1103/PhysRevD.41.3720;%%
C. Grosse-Knetter, \textit{Effective Lagrangians with higher derivatives and equations of motion}, Phys. Rev. D \textbf{49}, 6709--6719 (1994).
%


\end{thebibliography}

\end{document}